\documentclass[a4paper,11pt]{article}
\pdfoutput=1 

\usepackage{pst-all}	
\usepackage{physics}

\usepackage{jcappub} 

\usepackage[T1]{fontenc} 
\usepackage{cleveref}
\usepackage{tikz}
\usetikzlibrary{matrix, fit, positioning,arrows.meta,arrows, ,decorations.pathreplacing,shapes.geometric,backgrounds}

\title{\boldmath \texttt{CosmoMIA}: Cosmic Web-based redshift space halo distribution}

\author[1]{D. Forero Sánchez,\note{daniel.forerosanchez@epfl.ch}}
\author[2,3]{F.-S. Kitaura,\note{fkitaura@iac.es}}
\author[4,5,2,3]{F. Sinigaglia,}
\author[6,2]{J.M. Coloma-Nadal,}
\author[1]{J.-P. Kneib}

\affiliation[1]{Institute of Physics, Laboratory of Astrophysics, École Polytechnique Fédérale de Lausanne (EPFL), Observatoire de Sauverny, CH-1290 Versoix, Switzerland}
\affiliation[2]{Instituto de Astrof\'isica de Canarias, Calle Via L\'actea s/n, E-38205, La  Laguna, Tenerife, Spain}
\affiliation[3]{Departamento  de  Astrof\'isica, Universidad de La Laguna, E-38206, La Laguna, Tenerife, Spain}
\affiliation[4]{Département d’Astronomie, Université de Genève, Chemin Pegasi 51, CH-1290 Versoix, Switzerland}
\affiliation[5]{Institut für Astrophysik, Universität Zürich, Winterthurerstrasse 190, CH-8057 Zürich, Switzerland}
\affiliation[6]{Institute of Space Sciences (ICE, CSIC), Campus UAB, Carrer de Can Magrans, s/n, 08193 Barcelona, Spain}

\emailAdd{daniel.forerosanchez@epfl.ch}
\emailAdd{fkitaura@iac.es}

\abstract{Modern galaxy surveys demand extensive survey volumes and resolutions surpassing current dark matter-only simulations' capabilities. To address this, many methods employ effective bias models on the dark matter field to approximate object counts on a grid. However, realistic catalogs necessitate specific coordinates and velocities for a comprehensive understanding of the Universe.
In this research, we explore sub-grid modeling to create accurate catalogs, beginning with coarse grid number counts at resolutions of approximately $5.5\,h^{-1}\rm Mpc$ per side. These resolutions strike a balance between modeling nonlinear damping of baryon acoustic oscillations and facilitating large-volume simulations. Augmented Lagrangian Perturbation Theory (ALPT) is utilized to model the dark matter field and motions, replicating the clustering of a halo catalog derived from a massive simulation at $z=1.1$.
Our approach involves four key stages:
    Tracer Assignment: Allocating dark matter particles to tracers based on grid cell counts, generating additional particles to address discrepancies.
    Attractor Identification: Defining attractors based on particle cosmic web environments, acting as gravitational focal points.
    Tracer Collapse: Guiding tracers towards attractors, simulating structure collapse.
    Redshift Space Distortions: Introducing redshift space distortions to simulated catalogs using ALPT and a random dispersion term.
    Results demonstrate accurate reproduction of monopoles and quadrupoles up to wave numbers of approximately $k=0.6\,h$ Mpc$^{-1}$. This method holds significant promise for galaxy surveys like DESI, EUCLID, and LSST, enhancing our understanding of the cosmos across scales.}

\begin{document}
\maketitle
\flushbottom

\section{Introduction}
\label{sec:intro}

Large Scale Structure (LSS) analyses provide information on the content and evolution of the Universe. Over the last 20 years, various experiments have pushed the boundaries of observations exponentially. The pioneering Two Degree Field Galaxy Redshift Survey (2dFGRS) captured around 390,000 galaxy redshifts within a sky area of $2\times 10^3~\deg^2$ and a volume of $0.12~\mathrm{Gpc}^3/h^3$ \citep{Coless2001}. Following suit, the Baryon Oscillation Spectroscopic Survey (BOSS) under the Sloan Digital Sky Survey (SDSS) probed 1.5 million galaxies sprawled over an area of $\sim 10^4\deg^2$ \citep{Eisenstein2011, Dawson2013}, corresponding to a volume of around $2.5~\mathrm{Gpc}^3/h^3$. The extended BOSS (eBOSS) survey extended the depth of BOSS, which increased observed volume by $\sim1.5~\mathrm{Gpc}^3/h^3$. The Dark Energy Spectroscopic Survey (DESI) is expected to observe over 30 million redshifts in $14\times 10^3~\deg^2$ \citep{Levi2019, DESI2023a, DESI2023b} which amounts to a volume of $\sim20~\mathrm{Gpc}^3/h^3$. Simultaneously, the Euclid satellite \footnote{\url{http://www.euclid-ec.org}} is observing approximately 35 million galaxy redshifts and 1.5 billion shapes in order to combine galaxy clustering and weak lensing measurements \citep{Laureijs:2011gra}.

Additionally, future-generation surveys are on the horizon, including 4MOST\footnote{\url{http://www.4most.eu/}} (4-metre Multi-Object Spectroscopic Telescope) \citep{deJong2019}, HETDEX\footnote{\url{http://hetdex.org}}\citep{Hill:2008mv} (Hobby-Eberly Telescope Dark Energy Experiment), PFS\footnote{\url{https://pfs.ipmu.jp}}(Subaru Prime Focus Spectrograph) \citep{2014PASJ...66R...1T} and the Roman Space Telescope\footnote{\url{https://roman.gsfc.nasa.gov}}. These undertakings aim to decipher the properties of dark energy (DE) through meticulous measurements of the Baryon Acoustic Oscillations (BAO) scale and the universe's growth rate via Redshift Space Distortions (RSD). In essence, the forthcoming years are poised to witness a substantial amplification in the observable volume of the universe.

Uncertainties in the cosmological measurement are traditionally estimated using a sample of simulated universes that emulate the observed data. This requires, in particular, that these mocks have a volume at least as large as the observations, in order to properly estimate the statistical uncertainties in the data. However, current surveys such as DESI are already too big for an $N$-body simulation to be able to capture the total observed volume. The AbacusSummit \citep{Maksimova2021} simulation suite contains boxes of only $8~\mathrm{Gpc}^3/h^3$ which are not enough to contain the observed light cone. To encompass all tracers, including quasars, boxes with a volume of approximately $1000~\mathrm{Gpc}^3/h^3$ would be necessary. In addition, in order to estimate accurate covariances for cosmological measurements, it is necessary to perform thousands of simulations. Moreover, the number of simulations which are required  increases as the data vector size does, so the advent of multi-tracer and alternative analyses \citep{Zhao2022, Cuesta2023, Paillas2023} is expected increase the number of mocks to be produced.

To counter the huge computational burden of precise $N$-body simulations, fast and approximate simulations are performed. These simulations sacrifice both mass resolution and gravitational evolution accuracy in order to decrease computing time. Over the years, many such methods have emerged. Some of these use variations of Lagrangian perturbation theory in order to estimate the evolved dark matter (DM) density field \citep{Kitaura_2013, Kitaura_14,Chuang2015, Zhao2021}, whereas some others use particle mesh (PM) methods \citep{Tassev2013, Ding2023, Variu2023}. Regardless of the method, if the volume requirements of the survey are large enough, the resolution of the mocks must decrease in order to maintain the simulations computationally tractable, which means that only large scales will be appropriately modelled by the simulations. In order to take full advantage of observations, especially of small-scale information, our mocks and models should be able to accurately capture them, which low-resolution simulations alone cannot do. 

In this work we introduce the \textbf{Cosmo}logical \textbf{M}ultiscale \textbf{I}nfall \textbf{A}lgorithm (\texttt{CosmoMIA})\footnote{\url{https://github.com/dforero0896/CosmoMIA.jl}}; a model that corrects the positions of particles given by a low resolution approximate simulation in order to accurately emulate the subgrid clustering from a target $N$-body simulation, that is, the clustering on scales smaller than the spatial resolution of the approximate simulation. In \cref{sec:data} we briefly introduce the data used to showcase the capabilities of our model as well as including a short description the approximate gravity solver used in this work, then in \cref{sec:method} we fully describe our subgrid method. \Cref{sec:metrics} introduces the metrics used to evaluate and optimize the model parameters. In \cref{sec:results} we show our results and finally we discuss and conclude in \cref{sec:discussion}.

\section{Data}
\label{sec:data}
In order to apply our model, we need two main ingredients: a reference and a fast approximate simulation. For the former, we use an $N$-body simulation to extract the relevant target clustering on the scales desired, plus it allows to minimize the effects of cosmic variance through the use of its initial conditions in the approximate solver. On the other hand, the approximate simulation is the basis upon which we want to improve, correcting their small-scale clustering.
\subsection{$N$-body reference simulations}
To introduce our model in this particular case, we use halo catalogs from the publicly available \textsc{AbacusSummit} suite of simulations \citep{Maksimova2021} which are based on the \texttt{Abacus} code \citep{Garrison2019, Garrison2021}. The simulation suite consists of 97 different cosmologies around the central ``base'' Planck cosmology \citep{Planck2018}, each of which was run with a mass resolution of $2\times 10^9~\mathrm{M}_\odot /h$ ($6912^3$ particles in a $(2~\mathrm{Gpc}/h)^3$ volume) in order to meet the requirements of the DESI \citep{Levi2019} tracers. In this work, we use a subsample of one realization of the Abacus halo catalogs in the base cosmology, at a redshift of $z=1.1$. This downsampling is done such that the most massive halos are used until the number density of halos in the box is $3\times10^{-3}~(h / \mathrm{Mpc})^3$. This corresponds to $8\times10^6$ halos.

Furthermore, we use the corresponding initial conditions in order to run an approximate low-resolution simulation using the \texttt{WebON} code \cite{Kitaura2024}. The Abacus initial density fields, initially on a grid of $576^3$ cells, were downsampled with a sharp Fourier space filter to a mesh of $N_{\rm low} = 360^3$ cells, which corresponds to a cell side size of $\Delta x\approx5.5~\mathrm{Mpc}/h$. In terms of mass resolution, our approximate simulation has $\sim 10^{-4}\times$ less resolution than the reference.

We must emphasize the multiple reasons behind conducting this study at the chosen resolution. 
Firstly, a minimum resolution of between 5 and 10 $\rm Mpc$ is necessary to adequately resolve the displacement and peculiar velocity fields for accurately modeling the BAO  and the RSD \citep[see, e.g.,][]{2017MNRAS.467.2331V}.  Additionally, we aim at exploring  the lowest possible resolution to showcase the effectiveness of our method. 

Such low resolutions facilitate the efficient generation of large-volume mock datasets.
In fact, we conducted a series of 200 large-volume light-cone dark matter simulations using the \texttt{WebON} code within cubical volumes of $1000\,\mathrm{Gpc}^3/h^3$ \cite{Kitaura2024}, precisely with the resolution explored in this study.
\subsection{Approximate simulations}
\label{sec:gravsol}

A very precise description of the evolution of the ensemble of simulation particles under gravity can in general be computed with $N$-body codes such as \texttt{Abacus} \cite{Garrison2019}, \texttt{Gadget3} \citep{Springel2005} or \texttt{PkdGrav3} \citep{Potter2017}, given that enough particles are used i.e. there is a fine enough mass resolution. On top of the high resolution requirements, current and future surveys impose large volume requirements that are practically intractable to an $N$-body simulation. Even when these requirements are relaxed, these simulations remain too expensive to be used to estimate cosmological measurement covariances and thus estimate uncertainties in cosmological parameters. In order to estimate these, cosmological analyses have used fast simulations. These are simulations that are relatively cheaper in terms of compute time, but sacrifice mass resolution and accuracy in the gravitational evolution, usually because they use an approximate gravity solver, which models the real displacement field analytically or with a reduced-resolution particle-mesh (PM) solution. In this study, we use Augmented Lagrangian Perturbation Theory (ALPT) \citep{Kitaura_14}, an analytical approximation to the nonlinear displacement field that improves upon the second order LPT (2LPT) by modelling the small-scale displacement with spherical collapse (SC). This gravity solver was used in the Patchy mocks \citep{Kitaura_2013} used by BOSS. 

ALPT combines 2LPT for the long-range displacement and SC for small scales through a Gaussian kernel $\mathcal{G}(\vb*{q}, r_s)$ evaluated at the Lagrangian particle positions $\vb*{q}$ with a width of $r_s = 6~\mathrm{Mpc}/h$ (in this study). The ALPT displacement at redshift $z$ is then given by:
\begin{equation}
    \Psi_{\rm ALPT}(\vb*{q}, z) = \mathcal{G}(\vb*{q}, r_s)*\Psi_{\rm 2LPT}(\vb*{q}, z) + \left[\Psi_{\rm SC}(\vb*{q}, z) - \mathcal{G}(\vb*{q}, r_s)*\Psi_{\rm SC}(\vb*{q}, z)\right],
\end{equation}
where ``$*$'' denotes convolution.

The displacement field encodes all the necessary information for the application of our model. The final (Eulerian) DM particle positions are given by $\vb*{x} = \vb*{q} + \Psi_{\rm ALPT}$, whereas the Lagrangian coherent velocities are computed as
\begin{equation}
\vb*{v}_{\rm coh}(\vb*{q}, z) = \frac{H(z)}{1+z}\left(f_1(z)\Psi_{\rm Zeld} + f_2(z)\left[ \Psi_{\rm ALPT}(\vb*{q},z) - \Psi_{\rm Zeld}(\vb*{q},z)\right]\right)\,, 
\end{equation}
and the Eulerian velocities can be computed from these through $\vb*{v}_{\rm coh}(\vb*{x}, z) = \vb*{v}_{\rm coh}(\vb*{q} + \Psi_{\rm ALPT}(\vb*{q},z), z)$. The growth rates $f_{1,2}$ are given by \cite{Lahav1991}
\begin{align}
    f_1(z) = \Omega^{5/9}, \quad f_2(z) = \Omega^{6/11},\quad \Omega = \Omega_m\frac{(1+z)^3H_0^2}{H(z)^2},
\end{align}

Finally, we rely on the cosmological environment of the particles, not only in the use of the local dark matter density, but also in the form of a cosmic web type, which can be computed within the $\delta$-web prescription introduced in the companion papers \cite{Kitaura2024,Sinigaglia2021,Coloma2024}. This new cosmic web classification scheme identifies cells based on the eigenvalues $\lambda_\delta$ of the Hessian of the dark matter field $\Gamma_{ij} \equiv \partial_i\partial_j\delta$. Alternatively, the $\phi$-web prescription introduced by \citep{Hahn2007} may also be used. This classification relies on the eigenvalues $\lambda_\phi$ of the tidal field tensor $\mathcal{T}_{ij}\equiv\partial_i\partial_j\phi$, i.e. the Hessian of the potential. Both cosmic-web classification schemes require a threshold $\lambda_{\rm th}$ \cite{ForeroRomero2009} and define regions as
\begin{itemize}
    \item Knots: $\lambda_1, \lambda_2, \lambda_3 > \lambda_{\rm th}$
    \item Filaments: $\lambda_1, \lambda_2 > \lambda_{\rm th}\ \text{and}\ \lambda_3 < \lambda_{\rm th}$
    \item Sheets: $\lambda_1 > \lambda_{\rm th}\ \text{and}\ \lambda_2, \lambda_3 < \lambda_{\rm th}$
    \item Voids: $\lambda_1, \lambda_2, \lambda_3 < \lambda_{\rm th}$.
\end{itemize}
The present study, however, only uses the $\delta$-web classification. The inclusion of a hierarchical cosmic web classification (see \cite{Coloma2024}) in our subgrid modelling is left for future work. \Cref{tab:input-summary} summarizes the input information that our model requires.
\begin{table}
\centering
\caption{Summary of the input required for our method. The approximate quantities are expected to be generated by an approximate gravity solver.}
\label{tab:input-summary}
\begin{tabular}{l|l|l}
Required input & Description & Dimension \\ \hline
Target catalog & \begin{tabular}[c]{@{}l@{}}3D positions of all target particles for\\clustering measurements\end{tabular} & $(N_{\rm tracers},3)$ \\
Approx. Number counts & \begin{tabular}[c]{@{}l@{}}Number of target tracers per cell, normally\\the output of a bias model (i.e. Patchy).\end{tabular} & $(N_{\rm low})$ \\
Approx. DM positions & Final positions of DM particles & $(N_{\rm low},3)$ \\
Approx. DM velocity & Eulerian velocities of DM particles & $(N_{\rm low},3)$ \\
Approx. DM density & \begin{tabular}[c]{@{}l@{}}DM density painted on a mesh from DM\\particles\end{tabular} & $(N_{\rm low})$ \\
Approx. CW classification & \begin{tabular}[c]{@{}l@{}}Cosmic environment for each cell: voids,\\sheets, filaments or knots.\end{tabular} & $(N_{\rm low})$
\end{tabular}
\end{table}

In this work, we assume that a perfect bias model is available and our approximate number counts will be given by the real Abacus halo catalog painted into a mesh of $N_{\rm low}$ cells using Nearest Grid Point (NGP) mass assignment. 

\subsubsection{Velocity Kernel}

\label{sec:velcoh}


To accurately fit the halo velocities from the Abacus simulation, correcting the peculiar velocity field becomes imperative when approximate gravity solvers are utilized, as they frequently struggle to capture the nonlinear regime of structure formation effectively.

In the nonlinear regime, particles begin to undergo virialization, and the velocity field exhibits an additional dispersed component alongside the coherent one.
Handling the dispersed velocity component stochastically grants us a degree of freedom in adjusting the accuracy of the coherent component. The concept involves empowering the coherent velocity component with sufficient strength at smaller scales. 

To achieve this, we convolve the velocity field with an isotropic kernel designed to preserve power on large scales and inject more in the small scales. The enhanced coherent velocity field is given by
\begin{equation}
    \mathbf{v}_{\mathrm{coh}}'(k) = \mathbf{v}_{\mathrm{coh}}(k)\mathcal{K}(k) =\mathbf{v}_{\mathrm{coh}}(k) \left(1 + \frac{\Gamma}{1 + e^{-q}}k^2\right).
\end{equation}
\begin{figure}
    \centering
    \includegraphics[width=\linewidth]{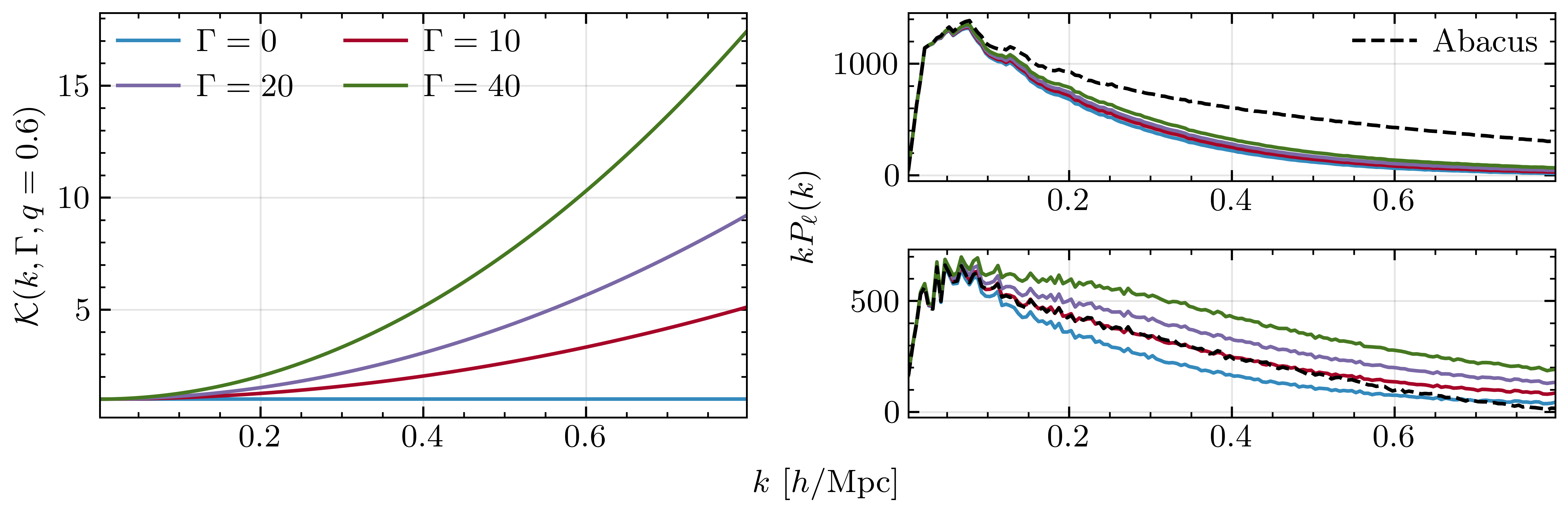}
    \caption{Velocity kernel used to increase the power of the approximate coherent velocity field. Left: Fourier-space representation of the kernel for different values of the gain parameter, $\Gamma$ -- including our fiducial choice $\Gamma=10$ -- and a fixed scale parameter $q=0.6$. Right: Effect of the velocity kernel on the monopole and quadrupole of the power spectrum of a sample with random subgrid assignment. We show the Abacus power spectrum as reference.}
    \label{fig:kernel}
\end{figure}
In principle, the gain and scale parameters of the kernel ($\Gamma,~q$ respectively) can be free. However, we find that the scale parameter does not significantly affect the two-point clustering when varying in the range of $q\in[0.2, 0.8]$ and thus fix it to $q=0.6$. The gain parameter, $\Gamma$ on the other hand, does significantly affect the clustering, specially the quadrupole, as can be seen in \cref{fig:kernel}. We find that a gain of $\Gamma = 10$ gives enough power on large scales for this particular application. These may vary from reference to reference.

Following this, at a later stage (see next section), we introduce a dispersed velocity component to systematically diminish this strength, thus enabling us to adjust the overall peculiar velocity on a global scale.

\section{Method}
\label{sec:method}

The method comprises four stages. 
Initially, we compute the dark matter density field on a mesh using approximate gravity solvers (see  \cref{sec:gravsol}). 
We allocate dark matter particle positions to the tracers with a novel adaptive  perturbative prescription (see  \cref{sec:particle}). Additionally, we compute the coherent peculiar velocity field using a nonlinear enhancement prescription (see  \cref{sec:velcoh}).
 The final and most novel aspect of the method, as outlined in  \cref{sec:subgrid}, entails assigning halo positions and peculiar velocities exploiting the phase-space and cosmic web information.

\subsection{Tracer Assignment}

\label{sec:particle}

Given the tracer number-counts field, to obtain mocks appropriate for LSS clustering it is necessary to obtain a discrete particle distribution. In order to do this, the appropriate number of particles are often generated randomly around the center of each cell, which ignores the existence of any substructure below the scale of the grid. As an alternative, the DM particles themselves can be regarded as a proxy for the tracers, which preserves some information on the sub-grid physics. However, there are often not enough particles to satisfy the requirements of the target tracer field, which usually implies that a larger (higher-resolution) simulation is required.

Our approach is to use a hybrid method. Let $N_t$ be the number of tracers required in each cell and $N_{\rm DM}$ the number of dark matter particles located in the same cell. Ideally, there are enough DM particles to assign to all tracers ($N_t \leq N_{\rm DM}$). However, there may be cases in which this condition is not satisfied. If $N_t > N_{\rm DM} > 0$, then the position of the $(N_{\rm DM}+l)-$th tracer is given by $\mathbf{x}_{N_{\rm DM}+l} = \mathbf{x}_{i \leq N_{\rm DM}} + \mathbf{G}$, that is, the new position is assumed to be the position ($\mathbf{x}_{i \leq N_{\rm DM}}$) of another particle in the cell plus a Gaussian-distributed random perturbation $\mathbf{G} \sim \mathcal{N}(\mu,\sigma)$, where $\mu=0$, $\sigma=0.1 \Delta\mathbf{x}$ and $\Delta\mathbf{x}$ is the cell size. Finally, for low resolution simulations, the case where $N_{\rm DM} = 0$ becomes more likely; in which case we sample the particle position randomly around the center of the cell, $\mathbf{c}$. The new particle position will then be given by $\mathbf{x}_{N_{\rm DM}+l} = \mathbf{c} + 0.5\Delta\mathbf{x}~\mathrm{sign}(\mathbf{U})(1-\sqrt{\abs{\mathbf{U}}})$ \citep{Chuang2015}, where $\mathbf{U}\sim\mathcal{U}(-1,1)$ is a random vector sampled from a uniform distribution between -1 and 1.

As a baseline, we will compare our method with catalogs generated from the same number-count field but using only randomly distributed particles as described in the previous paragraph. We refer to this catalog as the random sample.

In addition, during this step we assign coherent velocities $\mathbf{v}_{\mathrm{coh}}'$ to the new particles via Cloud-In-Cell (CIC) interpolation from the enhanced eulerian velocity fields (see \cref{sec:subgrid}). We have also tested assigning the cell's velocity (i.e., NGP mass assignment) but found that the velocity distribution in such case would not be centered around zero. Moreover, using the same CIC interpolation scheme we assign dark matter overdensity $\delta_{\mathrm{DM}}$. Finally, we use the NGP mass assignment scheme to assign a cosmic web type to each particle. These will be necessary in order to identify which  particles will behave as attractors in a given collapse iteration (see \cref{sec:subgrid}).

\subsection{Tracer Collapse}
\label{sec:subgrid}
\Cref{fig:power} shows (top-left panel) that randomly sampling particles within the cell will cause a loss of power with respect to the reference of 10\% at $k\approx0.1~h/\mathrm{Mpc}$ and 50\% already by $k\approx0.2~h/\mathrm{Mpc}$. In order to remedy this we perform a two-step particle collapse. In principle, the number of collapse steps is arbitrary, however we find that two steps are sufficient, preserve some computational efficiency, and are physically motivated.

This collapse consists on identifying pairs of particles, one of which will be the ``central'' whereas the other will be the ``satellite''. In the collapse, the satellite will be moved radially towards the central by a factor $\epsilon_c$. We do not collapse all particles towards all other particles, instead we use some criteria that allow us to decide whether a pair of particles is to be collapsed. These criteria correspond first to whether it the particle is a real DM particle from the simulation or has been sampled randomly in the cell; and second on their local cosmic web environment. Using these criteria we select the subset of particles that 1. are real DM and 2. do not reside in cosmic voids. These are denominated \textit{attractors}, and reflect the fact that we rely more on the DM particle positions than the randomly sampled ones and that we expect gravity to cause particles to collapse towards peaks in the DM field.

The first collapse step is done among these attractors exclusively, i.e., both central and satellite populations are the attractors and the selection criteria for collapsing pairs is simply given by the closest minimum distance. This means that we collapse attractors to their closest attractor. The second step is done taking the central population to be the attractors and the satellite population to be the rest. In this case, we modify the selection criteria to also include information on the local DM density. In particular, we will collapse satellites either to the closest central or to the central located in the highest density environment.
\subsection{Redshift Space Distortions}
The collapse step is isotropic and completely independent on tracer velocities. In order to generate accurate mocks, we must provide accurate redshift-space clustering, which requires us to tune the tracer velocities. In \cref{sec:velcoh} we introduced the isotropic kernel that we employ to enhance the coherent flow that the approximate gravity solver provides. This kernel has two parameters that depend on the reference but need not be fine tuned, that is, we only require the quadrupole (of the redshift-space clustering) to have enough power on small scales such that it can later be fine-tuned. This subsection explains how this fine-tuning is performed.

Once the quadrupole of the redshift-space clustering has enough power (i.e. more than the reference) on all scales, we add a dispersive velocity component to the satellite particles in order to account for tracer virialisation. The final form of the velocity dispersion is 
\begin{equation}
    \vb{v}_{\rm disp} = 10\vb{G'}\gamma\sqrt{1 + \hat{\delta}_{\rm DM}},
\end{equation}
where $\vb{G'}$ is a Standard-Gaussian-distributed random vector, $\hat{\delta}_{\rm DM}$ is the DM density of the central particle, clipped to the $[0,\infty)$ range and $\gamma$ is the free velocity dispersion parameter.
\subsection{Summary of the model}
Let us now summarize the collapse step, $i$:
\begin{enumerate}
    \item \textbf{Neighbour identification}: we limit the radius $r_{\mathrm{c},i}$ of influence of each collapse, which allows us to fine-tune small scales and be more computationally efficient. These searches are efficiently implemented in Julia using the \texttt{CellListMap.jl} package \citep{martinez_celllistmapjl_2022}.
    \item \textbf{Pair identification}: we define which particles will be collapsed (``satellites'') and towards which other particle (``central'') they will move. We do so based on their cosmic web environment, their radial distance and whether they were gravitationally evolved during the simulation or were randomly generated within the cell. We stress that the set of centrals/satellites is not constant and different subsets of particles can be used in each collapse iteration as explained before.
    \item \textbf{Collapse}: this steps effectively multiplies the radial distance between the satellites and the corresponding central $s$ by a factor $\epsilon_{\mathrm{c},i}$, that is, for $\epsilon_{\mathrm{c},i}<1$ the pair collapses.
    \item \textbf{Velocity dispersion}: we modify the velocity of the satellite particles by adding a stochastic or dispersive term such that their final velocity is $\vb{v}_{\rm sat} = \vb{v}'_{\rm coh} + \vb{v}_{\rm disp}$.
\end{enumerate}

A summary of the free parameters of our collapse model can be found on \cref{tab:model-summary}.
\begin{table}
\centering
\caption{Summary of the collapse model parameters, their description and their final values for our test on Abacus halos at $z=1.1$.}
\label{tab:model-summary}
\begin{tabular}{c|l|l}
\multicolumn{1}{l|}{Parameter name} & Description & Value \\ \hline
$r_{\mathrm{c}, 1}$ & \begin{tabular}[c]{@{}l@{}}Cutoff radius for nearest neighbor search during \\ attractor-attractor collapse\end{tabular} & 8.2 \\
$r_{\mathrm{c}, 2}$ & \begin{tabular}[c]{@{}l@{}}Cutoff radius for nearest neighbor/density peak \\ search during attractor-not attractor collapse\end{tabular} & 2.13 \\
$\epsilon_{\mathrm{c}, 1}$ & Collapse factor for attractor-attractor collapse & 0.73 \\
$\epsilon_{\mathrm{c}, 2}$ & Collapse factor for attractor-not attractor collapse & 1.77 \\
$\gamma~[\mathrm{km}/\mathrm{s}]$ & Velocity dispersion parameter & 4.75
\end{tabular}
\end{table}

\section{Evaluation metrics: Clustering statistics}
\label{sec:metrics}
In this section, we evaluate the performance of the method using several summary statistics. We begin by examining the two-point correlation function  in real and redshift space, both in Fourier and configuration space,  including multipole expansions (see \cref{sec:2PCF}).
Following that, we present our results on the three-point statistics in Fourier space, specifically focusing on the bispectrum (see \cref{sec:3PCF}). 

\subsection{Two-point clustering}
\label{sec:2PCF}

The two-point functions contain all the information of a Gaussian field. They are widely used in cosmology in order to extract information from the large-scale structure of the Universe \citep{Alam2017, Alam2021} and are the main observables of modern spectroscopic instruments such as DESI. Our main objective is to provide two-point statistics that are accurate in scales smaller than the approximate simulation's grid would allow. Given a catalog, we use the Fourier-space two-point function (i.e. the power spectrum).
\begin{equation}
    \label{eq:power-spectrum}
    (2\pi)^3\delta^{D}(\vb*{k}+\vb*{k'})P(k) = \langle\delta(\vb*{k})\delta(\vb*{k'})\rangle,
\end{equation}
where $\delta^{D}$ is a Dirac delta distribution, $\vb*{k}$ and $\vb*{k'}$ are wavevectors, $\delta(\vb*{k})$ is the Fourier-space overdensity field and $\langle\,\cdot\,\rangle$ denotes averaging. Through this work we estimate power spectra using the \texttt{CosmoCorr.jl}\footnote{\url{https://github.com/dforero0896/CosmoCorr.jl.git}} and \texttt{pypowspec}\footnote{\url{https://github.com/dforero0896/pypowspec.git}} codes, both based on the C \texttt{powspec}\footnote{\url{https://github.com/cheng-zhao/powspec.git}} code. 

However, spectroscopic observations cannot disentangle the cosmological redshift from the redshift due to the peculiar velocity of the tracers. in order to compute the redshift space clustering, we add the contribution of the peculiar velocity to the position of each tracer in the plane-parallel approximation, assuming a line of sight along the $Z$ (third) axis:
\begin{equation}
    \label{eq:rsd}
    Z_{z} = Z_r + v\frac{1+z}{H(z)},
\end{equation}

where $Z_z$ denotes the redshift space position along the line of sight, $v$ is the halo's peculiar velocity along the same direction, and $H(z)$ is the Hubble parameter at redshift $z$.

The peculiar velocity contributions induce anisotropy in the power spectrum which can be easily detected in its multipoles, $P_\ell(k)$. These are the projections of the 2-dimensional power spectrum $P(k,\mu)$ onto a Legendre polynomial basis $L_\ell(\mu)$, where $\mu$ is the cosine of the angle to the line of sight, that is
\begin{equation}
    \label{eq:power-multipoles}
    P_\ell(k) = \frac{2\ell+1}{2}\int_{-1}^1\dd\mu~P(k, \mu)L_\ell(\mu).
\end{equation}

Alternatively, in configuration space we compute the two-point correlation function $\xi(s)$. In practice, we estimate it using the natural estimator \citep{Pebbles1974}, which allows us to compute very small scale clustering that is impractical to compute in Fourier space:
\begin{equation}
    \label{eq:tpcf}
    \xi(s,\mu) = \frac{{\rm DD}(s,\mu) - {\rm RR}(s,\mu)}{{\rm RR}(s,\mu)}.
\end{equation}
$\mathrm{DD}(s,\mu)$ is the number of pairs separated by a distance $s$ in the data catalogue, normalised by the total number of pairs $N_D(N_D - 1)$, where $N_D$ is the number of tracers in the catalogue. Equivalently, the $\mathrm{RR}(s,\mu)$ term is the number of such pairs in a random catalogue. Given that our tests are performed on periodic boxes, the $\mathrm{RR}$ factor is computed analytically.

The multipoles of the correlation function are analogously defined as
\begin{equation}
    \label{eq:tpcf-multipoles}
    \xi_\ell(s) = \frac{2\ell+1}{2}\int_{-1}^1\dd\mu~\xi(s, \mu)L_\ell(\mu).
\end{equation}
Through this paper we use the correlation function computation implementations in the \texttt{CosmoCorr.jl} and \texttt{pyfcfc}\footnote{\url{https://github.com/dforero0896/pyfcfc.git}}, the latter of which is based on the C \texttt{FCFC}\footnote{\url{https://github.com/cheng-zhao/FCFC.git}}\citep{Zhao2021, Zhao2023} code.
\subsection{Bispectrum}
\label{sec:3PCF}

While the two-point clustering is a good descriptor of the matter field, higher order statistics such as the bispectrum are being increasingly used in order to constrain cosmology \citep{Sefusatti2006, Gil2017, Sugiyama2023a} and break degeneracies present in the two-point analyses. Moreover, these higher order statistics are instrumental in constraining alternative cosmological models including modified gravity \citep{Borisov2009, Gil2011} and primordial non-Gaussianity \citep{Gangui1994, Sefusatti2007, Welling2016}. Current and next generation mocks should then be able to properly emulate three-point functions and their covariances. 

The bispectrum is defined as
\begin{equation}
    \label{eq:bispectrum}
    \delta^{D}(\vb*{k}_1+\vb*{k}_2+\vb*{k}_3)B(k_1, k_2, k_3) = \langle\delta(\vb*{k}_1)\delta(\vb*{k}_2)\delta(\vb*{k}_3)\rangle.
\end{equation}
In this work we evaluate the bispectrum in a configuration with $k_1 = 0.1$ and $k_2 = nk_1$ with $n=2,~3$. $k_3$ is given by the closure relation enforced by the Dirac distribution $\delta^D$. For this bispectrum projection, we show it as a function of the angle $\theta_{12}$ between the wavevectors $\vb*{k}_1$ and $\vb*{k}_2$, given by $k_3^{2} = k_2^2\sin^2\theta_{12} + (k_2\cos\theta_{12} + k_1)^2$. In practice, we use the GPU-enabled bispectrum implementation available in the \texttt{jax-powspec}\footnote{\url{https://github.com/dforero0896/jax-powspec.git}} package, based on \texttt{Pylians3}\footnote{\url{https://github.com/franciscovillaescusa/Pylians3.git}} and a Julia wrapper to the C library \texttt{libbispec}.

Moreover, recent analyses \citep{Sugiyama2021, Sugiyama2023a, Sugiyama2023b}, have used the Sugiyama estimator \cite{Sugiyama2019} of the bispectrum in order to constrain cosmology. This estimator computes the bispectra in different $k_1,~k_2$ configurations for different multipoles $\ell_1,~\ell_2,~L$. A common configuation is the diagonal $k_1=k_2\equiv k$ configuration. \citep{Sugiyama2019} shows that the $\ell_1 = \ell_2 = L = 0$ and $\ell_1 = \ell_2 = 2,~L = 0$ have a high signal to noise ratio and are thus important for cosmological analyses. In practice we use the \texttt{Triumvirate} package \citep{Wang2023} to compute the bispectra.

\subsection{Free Parameters \& Optimization}
Our model has just 5 free parameters, $\theta$, that can be optimized by trial and error to a certain extent. Our approach however is to use a minimizer to compute steps towards a reasonable-looking minimum. To do so, we define the loss function as the weighted sum of the mean absolute errors, $\mathrm{MAE}(\theta) = \frac{1}{N}\sum_{i=0}^N\abs{x_i - \hat{x}_i(\theta)}$ of different predicted clustering statistics $\hat{x}(\theta)$ compared to the target statistics $x_i$. Our total loss is then
\begin{equation}
    \mathcal{L} = a_1\mathcal{L}_{P} + a_2\mathcal{L}_{P,2}, a_3\mathcal{L}_{P,0} + a_4\mathcal{L}_{\xi,0},
\end{equation}
where $\mathcal{L}_P$ is the MAE of the real-space power spectrum monopole, $\mathcal{L}_{P,\ell}$ is the redshift space power spectrum $\ell-$multipole and $\mathcal{L}_{\xi,0}$ is the redshift space correlation function monopole. The weighting factors $a_i$ not only help to regularize the dynamic range of the different statistics, but modifying them during the training of the model will help to speed up convergence. In the first stages of training we focus on the power spectrum in redshift space, making sure that $a_2>a_3>a_1$ and $a_4=0$. Once the larger scales have been fixed, we introduce the correlation function term with $a_4=50$ in order to help the model better fit the smallest scales. In practice we use the \texttt{Optim.jl}\footnote{\url{https://github.com/JuliaNLSolvers/Optim.jl.git}} package.
\section{Results}
\label{sec:results}
\begin{figure}
    \centering
    \includegraphics[width=\linewidth]{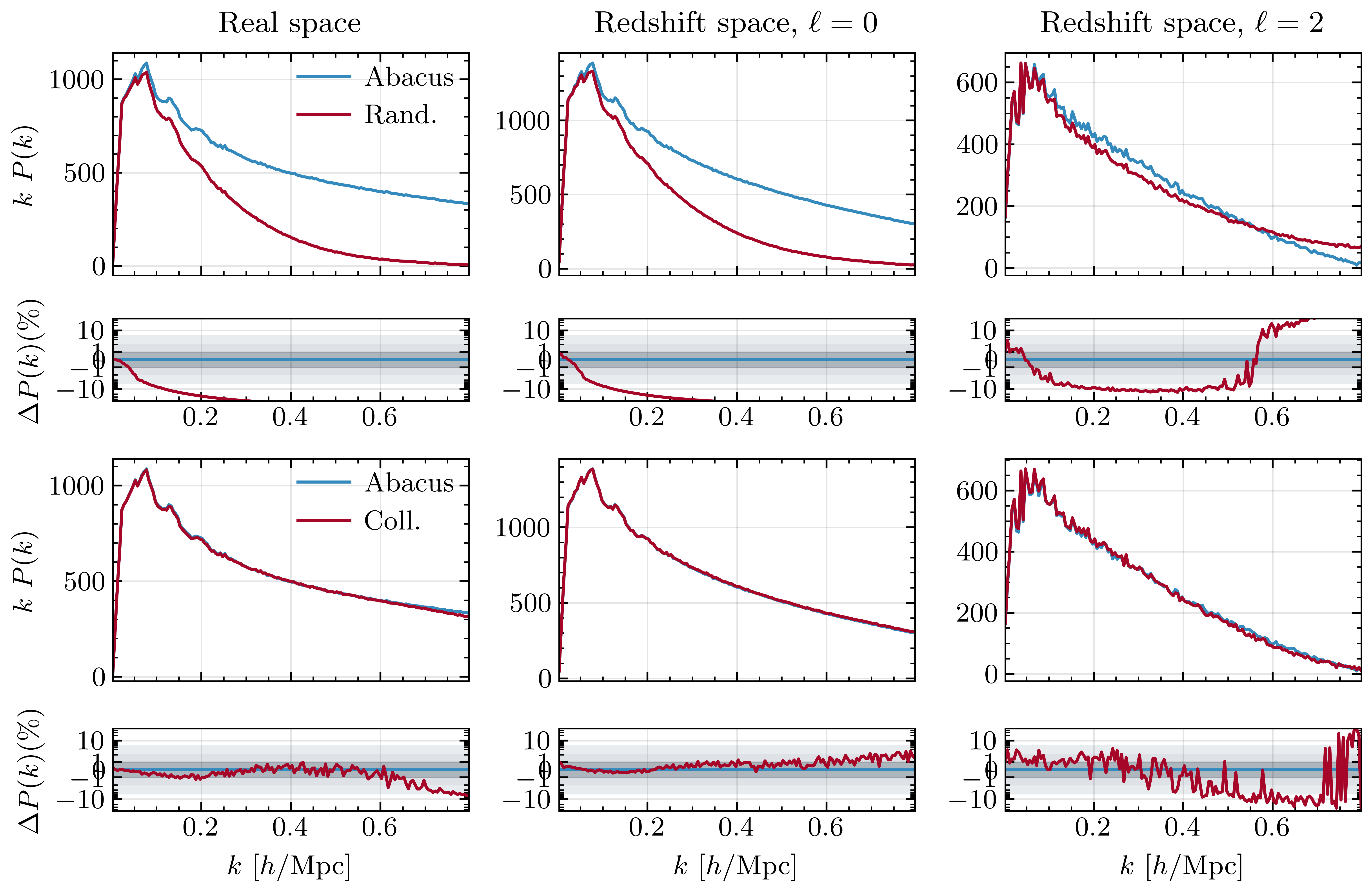}
    \caption{Power spectra of the reference Abacus halo catalog, a randomly sampled population of halos and our collapsed sample. Left: Real space monopole, : Redshift space monopole. Right: Redshift space quadrupole. Top row: comparison of the Abacus reference to a random sample. Bottom: Comparison of Abacus against our collapsed sample. Monopoles agree to 1\% (dark grey band) up to $k\sim 0.6 h/{\rm Mpc}$. The quadrupole agrees within 5\% (light grey band) on the same scale range.}
    \label{fig:power}
\end{figure}
Our fitting procedure can be run for an arbitrary amount of time, however after a couple of hours the precision is already good enough compared to the reference. This usually means that the redshift-space power spectrum has reached $1-2\%$ accuracy down to scales of $0.4-0.5~h/\rm Mpc$. Notice that given that we share the same initial conditions, we do not take cosmic variance into account for the fit. The resulting values of our fit are also shown in \cref{tab:model-summary}. The values of the collapse fractions are interesting since the first collapse seems to be reducing the pairwise distances to about 70\% of the original value, this will generate excess power at scales under the first cutoff radius. The subsequent ``collapse'' step actually separates particles from the chosen attractors, diluting power in the very small scales, under the second collapse radius. The combination of these two accurately emulates the small-scale clustering. We comfortably recover the $~50\%$ loss in power at large scales and inject the necessary power to scales significantly smaller than the minimum requirements for BAO and RSD analyses.

\Cref{fig:power} shows the Fourier-space two-point clustering of the reference Abacus halo catalog compared to the baseline random subgrid particle assignment and our collapsed model for the subgrid clustering. The leftmost column shows that the real-space clustering is within 1\% of the target up to scales of $k\approx0.6~h/\mathrm{Mpc}$. This is a substantial improvement over the usual particle distribution technique where at these scales the power is already negligible. However, our main target is the two-point redshift space clustering, which we upweight in the loss function. The monopole is consistent with the reference to 1\% precision up to $k\approx0.5~h/\mathrm{Mpc}$ and 2\% up to $k\approx0.7~h/\mathrm{Mpc}$. The quadrupole shows also a significant improvement, as it is consistent within 5\% up to scales of $k\approx0.4~h/\mathrm{Mpc}$.

\begin{figure}
    \centering
    \includegraphics[width=\linewidth]{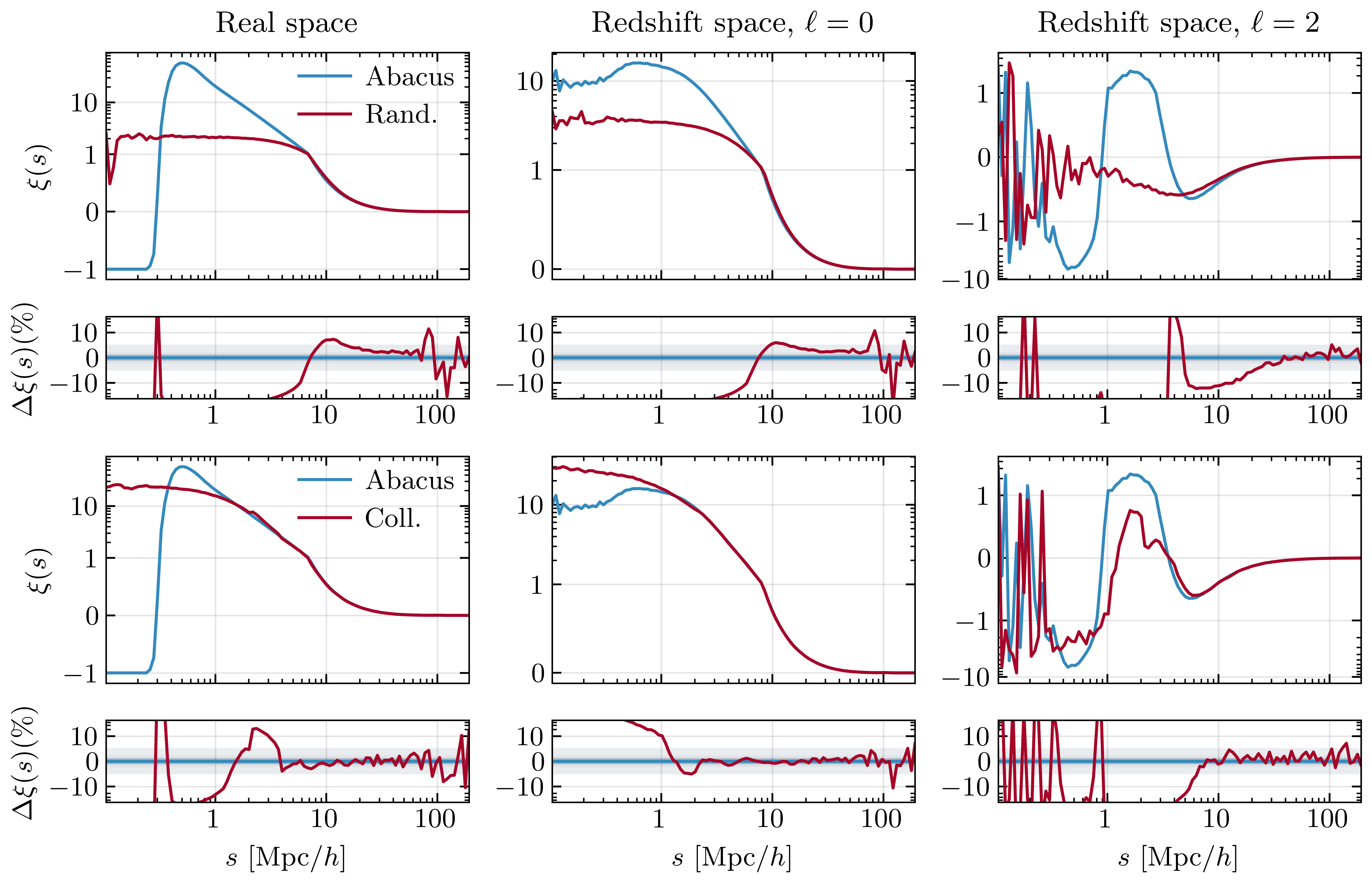}
    \caption{Small-scale correlation function of the reference Abacus halo catalog, a randomly sampled population of halos and our collapsed sample. Left: Real space monopole, : Redshift space monopole. Right: Redshift space quadrupole. Top row: comparison of the Abacus reference to a random sample. Bottom: Comparison of Abacus against our collapsed sample. Monopoles agree to 1\% (grey band) down to to $s\sim 3 {\rm Mpc}/h$. The quadrupole agrees within 5\% down to to $s\sim 10 {\rm Mpc}/h$. The zero-crossing of the quadrupoles makes the ratio diverge.}
    \label{fig:tpcf}
\end{figure}

\begin{figure}
    \centering
    \includegraphics[width=\linewidth]{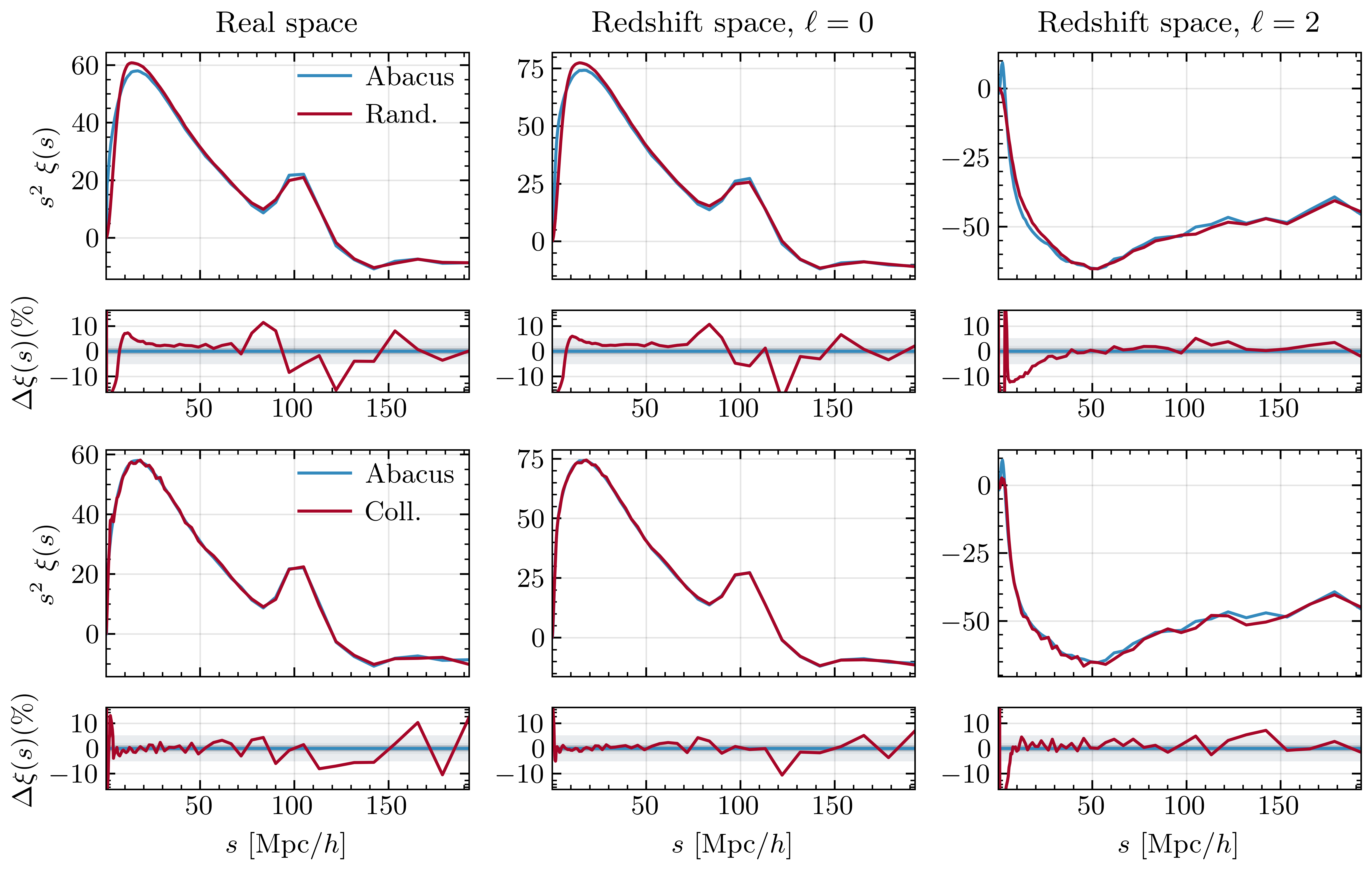}
    \caption{Large-scale correlation function of the reference Abacus halo catalog, a randomly sampled population of halos and our collapsed sample.}
    \label{fig:tpcf-ls}
\end{figure}

The late addition of the configuration space monopole to the loss, allows for fine-tuning the smallest scales in order to properly reproduce the redshift space monopole. \cref{fig:tpcf} shows that the term used in the loss (redshift space monopole) is accurate within 2\% down to scales of $s\approx5~\mathrm{Mpc}/h$ and 5\% down to $s\approx1~\mathrm{Mpc}/h$. On the other hand, the real-space case does not fit as well, specially on these small scales. The feature at $s\approx0.5~\mathrm{Mpc}/h$, likely due to halo exclusion, makes the fit to these scales considerably more difficult, and the model was not explicitly trained on this particular measurement. Notice how the redshift space mapping, largely mitigates these effects and the model is able to capture the small scales better. We expect that this feature will not be present in galaxy clustering, thus improving the overall fit to these scales. Nonetheless, we find a 5\% agreement with the reference down to $s\approx4~\mathrm{Mpc}/h$ in the real space two-point function. Similarly the quadrupole is also not explicitly trained on thus we do not capture the feature at $s\approx2~\mathrm{Mpc}/h$. However, we are able to recover the agreement with the reference even at $s\approx5~\mathrm{Mpc}/h$. This is also a significant improvement over the baseline, which in this case does not contain a dispersion velocity term. Finally, \cref{fig:tpcf-ls} shows that the random assignment of particles performed in the baseline, dampens the strength of the BAO peak significantly. This was already seen for the EZmocks \citep{Chuang2015}, where the artificial strengthening of the BAO peak in the initial conditions was necessary. Evidently, our model not only allows to fix small-scale clustering  but also corrects smaller effects on relevant larger scales such as the sound horizon.

\begin{figure}
    \centering
    \includegraphics[width=\linewidth]{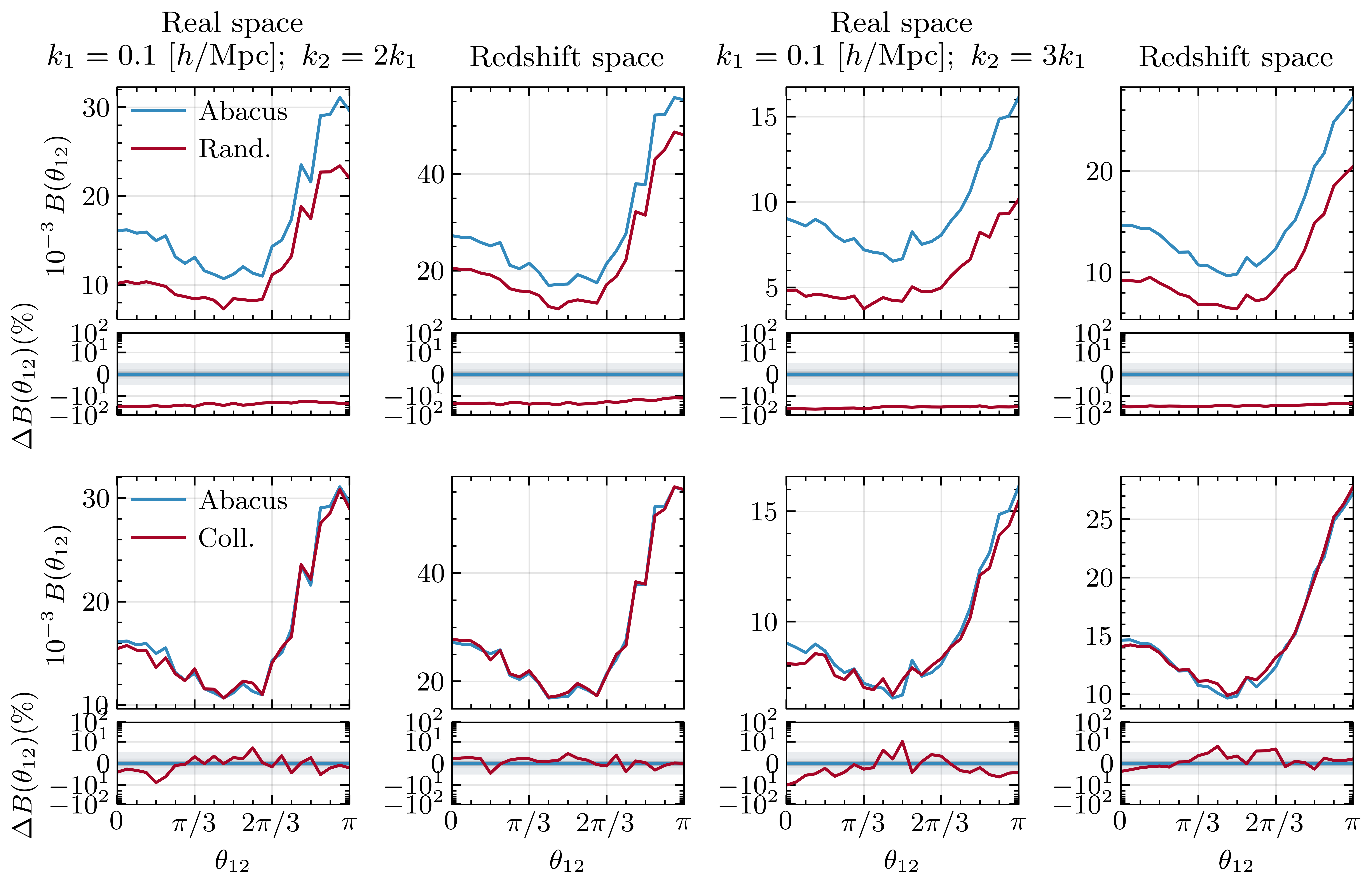}
    \caption{Bispectrum of the reference Abacus halo catalog, a randomly sampled population of halos and our collapsed sample for two different triangle configurations. Left two columns: Real and redshift space curves for $k_1 = 0.1~h/\mathrm{Mpc};~k_2=2k_1$. Right columns: Real and redshift space curves for $k_1 = 0.1~h/\mathrm{Mpc};~k_2=3k_1$. Top row: comparison of the Abacus reference to a random sample. Bottom: Comparison of Abacus against our collapsed sample. Gray areas show 5\% discrepancy. The $k_2=2k_1$ agrees with the reference within 5\% while the $k_2=3k_1$ configurations agree within 10\%.}
    \label{fig:bispec}
\end{figure}

\begin{figure}
    \centering
    \includegraphics[width=\linewidth]{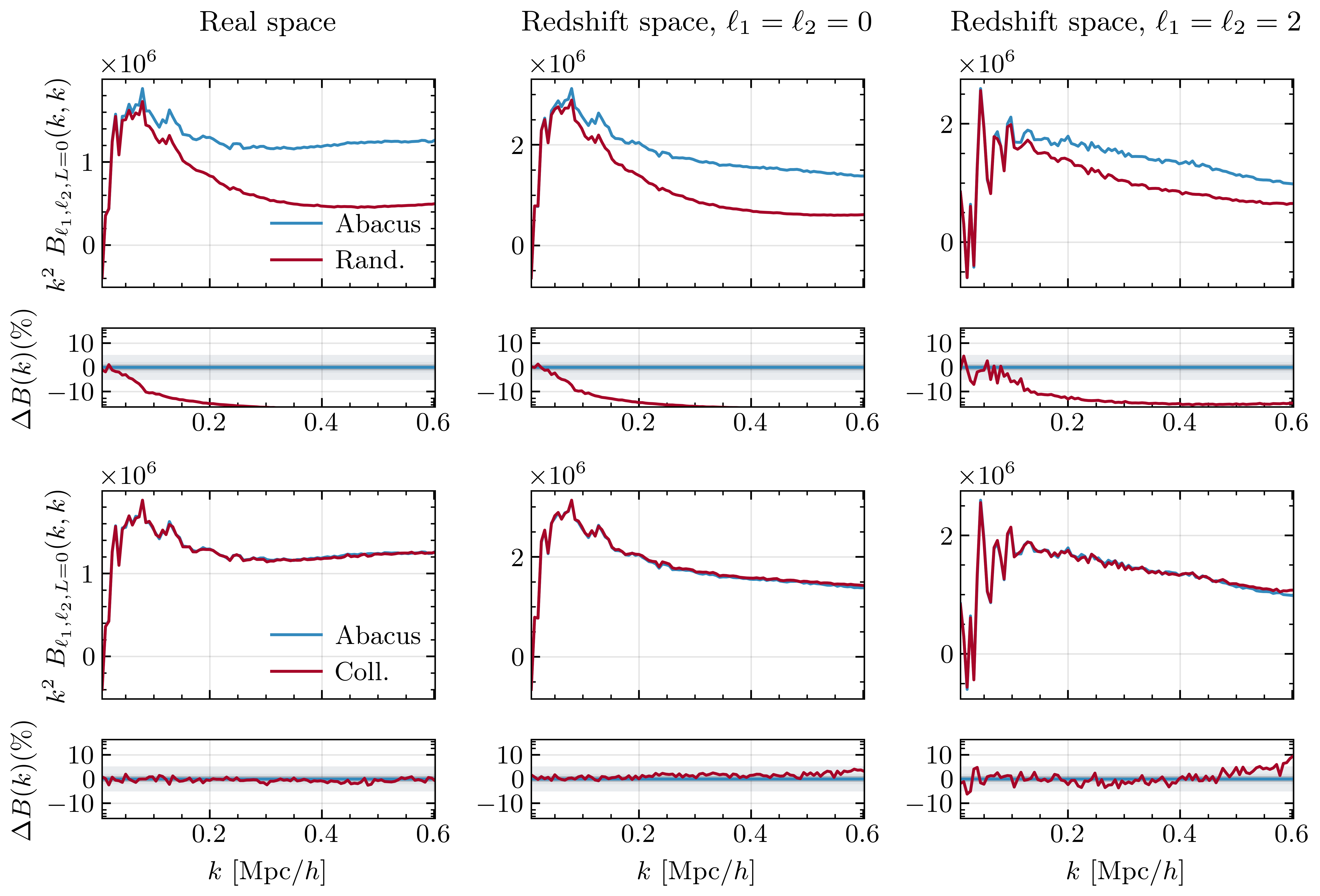}
    \caption{Diagonal $B_{\ell_1\ell_2 L}(k,k)$ configuration of the bispectrum in different $(\ell_1,\ell_2,L)$ configurations of the Sugiyama estimator as a function of $k$. We show two of the highest signal-to-noise configurations. Our collapsed catalog matches the reference within 5\% up to $k = 0.5~h/\mathrm{Mpc}$ in both real and redshift space for both configurations shown.}
    \label{fig:bk}
\end{figure}

In order to model the covariance correctly, it is necessary to properly model the three-point clustering of the sample \citep{Baumgarten2018}. In \cref{fig:bispec} we show the real and redshift space for one large-scale $k_1=0.1~h/\mathrm{Mpc},~k_2=2k_1$ and one smaller scale $k_1=0.1~h/\mathrm{Mpc},~k_2=3k_1$ $B(\theta_{12})$ configurations. Evidently, the baseline particle assignment does not generate accurate bispectra, showing discrepancies of around 50 and 100\% for the large and small scale distributions respectively. Given that we are using ideal number counts, this also shows that properly fitting the global PDF of number counts with a bias model is not enough for an accurate bispectrum and that an accurate subgrid description is necessary. On the other hand, our collapsed sample shows a significantly better agreement with the reference, with per cent differences of 5 and 10\% for the large and small scale bispectrum configurations in redshift space, while the agreement for the real-space clustering is slightly worse. This is not surprising given that we focused on fitting the redshift space measurements.

Furthermore, we want to check whether our model can reproduce some of the common bispectrum projections used for cosmological measurements. In order to do this, we use the Sugiyama estimator \citep{Sugiyama2019} and choose two configurations with a high signal-to-noise ratio, thus the most relevant for cosmological parameter estimation. We use the $\ell_1=\ell_2\L=0$ and $\ell_1=\ell_2=2,~L=0$ configurations. For the former we show the real and redshift space and for the latter we show the redshift space only measurements in \cref{fig:bk}. We observe that the random subgrid assignment causes the bispectrum multipoles shown to have a divergence of 50\% compared to the reference at scales as large as $k=0.2~h/\mathrm{Mpc}$. On the contrary, our collapse method can match the Abacus 3-point clustering within 5\% down to scales of $k\approx0.6~h/\mathrm{Mpc}$. This kind of agreement may open the door for more precise cosmological analyses provided that theoretical models can also reach these scales.

\section{Discussion and Conclusions}
\label{sec:discussion}
The present work presents a subgrid model designed to emulate the small-scale clustering of an accurate ($N$-body) reference using an approximate, fast simulation with a mass resolution of $10^{-4}$ times the original one. We have explored not only the technical details of the modelling of the small-scale clustering but have also identified theoretically motivated solutions to offset the limitations of low resolution gravity solvers. 
We compare our model with a baseline created using a common (random) particle assignment scheme within each cell.

We show that even under the assumption that a perfect bias model exists, such that the target number counts field is exactly the one obtained from the simulation, the random particle assignment is not enough to obtain two and three point statistics that properly emulate the reference on the relevant scales. Within our model, we employ a dark matter-random hybrid particle position assignment within each low resolution simulation cell that potentially mitigates the dampening of the baryon acoustic peak that is observed in other methods such as EZmocks \citep{Chuang2015}. In addition, we developed a two-step collapse and disperse velocity model with 5 free parameters that takes into account cosmic-web and environmental information and can automatically be tuned to reproduce the real and redshift space two point clustering of the reference down to scales of $k\approx0.6~h/\mathrm{Mpc}$ or $s\approx1~\mathrm{Mpc}/h$ with an accuracy of 1\% in the monopole and of 5\% in the quadrupole. This accuracy is a large improvement compared to previous approaches to approximate mock generation as seen in \cite{Chuang2015b}, where only the $N$-body based methods such as COLA yield comparable results. This agreement satisfies the requirements from current generation large spectroscopic instruments.

We employ a multi-stage training scheme that allows for faster optimization by avoiding the computation of sub-dominant terms in the objective function such as the small scale configuration space two-point clustering. This term will only be computed in later stages of the training, when large scales have already been fit by minimizing the power spectrum terms of the loss.

Despite not being trained to do so, our model is also capable of fitting various bispectra configurations $B(\theta_{12})$ with accuracies of 5 to 10\%, a factor of 10 improvement over the baseline random particle assignment. This hints at the capability of our model to properly model the two-point covariance matrix, as shown by \citep{Baumgarten2018}. Moreover, we test different cosmologically relevant bispectrum projections using the Sugiyama estimator. We find a 5\% agreement with the reference, which is also a factor of 10 improvement over the naive random assignment of particles. The combination of these shows that this novel technique can be applied to the generation of mocks for higher order statistics analyses, which will be a major focus of spectroscopic surveys in the coming years.

There are several avenues for enhancing our method further. While in this study we have employed ALPT at extremely low resolutions (four orders of magnitude less than the reference simulation), alternative approaches such as eALPT \citep[][]{2023arXiv230103648K} operating at higher resolutions or machine learning-based solutions \citep[][]{2019PNAS..11613825H} offer promising alternatives. These methods could provide an improved baseline for determining the positions and velocities of dark matter particles.
The particle collapse step could potentially be enhanced using a machine learning approach. However, in this work, we aimed to demonstrate how we can leverage the information contained in the phase-space distribution of particles from approximate gravity solvers through straightforward prescriptions without requiring large training data sets.

Our model can be regarded as a post-processing step that is completely independent of the method used to model the tracer field, thus various bias models such as \citep{Kitaura_14, Chuang2015, Schmittfull2019, Schmittfull2021} can equivalently be used. In particular, this allows us to overcome the data-volume limitations of machine learning (ML)-enabled models or models that are in general limited by (GPU) memory such as \citep{Li2022}, given that a lower resolution mesh is required. In addition, this step can be applied to existing mocks such as the MD-Patchy mock suite used in BOSS in order to mitigate the small-scale discrepancies with respect to $N$-body (e.g.,  \citep{Ereza2023}).

Finally, we stress that a great advantage of the model is its flexibility in the fact that modifications can easily be made to include more cosmic-web related information in the definition of attractors or even the number of collapse steps showing its potential paths for further improvement. This method will potentially become crucial in the analysis of current and upcoming galaxy surveys.

\section*{Acknowledgements}

DFS and JPK acknowledge support from the Swiss National Science Foundation (SNF) "Cosmology with 3D Maps of the Universe" research grant, 200020\_175751 and 200020\_207379. FSK, JMCN and FS acknowledge the Spanish Ministry of Economy and Competitiveness (MINECO) for financing the \texttt{Big Data of the Cosmic Web} project: PID2020-120612GB-I00 under which this work has been conceived and carried out, and the IAC for support to the \texttt{Cosmology with LSS probes} project. DFS thanks hospitality at the IAC.

\bibliographystyle{JHEP} 
\bibliography{references} 
\end{document}